\documentclass[10pt,conference]{IEEEtran}
\IEEEoverridecommandlockouts
\usepackage{cite}
\usepackage{amsmath,amssymb,amsfonts}
\usepackage{algorithmic}
\usepackage{graphicx}
\usepackage{textcomp}
\usepackage{xcolor}
\usepackage{multirow}
\usepackage{listings}
\usepackage{booktabs}
\usepackage{soul}
\usepackage{enumitem}
\usepackage[framemethod=tikz]{mdframed}
\usepackage{tikz}
\usepackage[hidelinks]{hyperref}

\def\BibTeX{{\rm B\kern-.05em{\sc i\kern-.025em b}\kern-.08em
    T\kern-.1667em\lower.7ex\hbox{E}\kern-.125emX}}

\makeatletter % changes the catcode of @ to 11
\newcommand{\linebreakand}{%
  \end{@IEEEauthorhalign}
  \hfill\mbox{}\par
  \mbox{}\hfill\begin{@IEEEauthorhalign}
}
\makeatother % changes the catcode of @ back to 12

\definecolor{valcolor}{rgb}{0.414,0.527,0.348}
\definecolor{backcolor}{rgb}{0.168,0.168,0.168}
\definecolor{keycolor}{rgb}{0.906,0.746,0.414}
\definecolor{attrcolor}{rgb}{0.727,0.727,0.727}
\definecolor{comcolor}{rgb}{0.5,0.5,0.5}

\makeatletter
\lst@CCPutMacro
    \lst@ProcessOther{"3C}{\lst@ttfamily{\textcolor{keycolor}{<}}{\textcolor{keycolor}{<}}}
    \lst@ProcessOther{"3E}{\lst@ttfamily{\textcolor{keycolor}{/> }}{\textcolor{keycolor}{/>}}}
    \lst@ProcessOther{"2A}{\lst@ttfamily{\textcolor{keycolor}{>}}{\textcolor{keycolor}{>}}}
    \lst@ProcessOther{"26}{\lst@ttfamily{\textcolor{keycolor}{</ }}{\textcolor{keycolor}{</ }}}
    \lst@ProcessOther{"3D}{\lst@ttfamily{\textcolor{valcolor}{=}}{\textcolor{valcolor}{=}}}
    \lst@ProcessOther{"21}{\lst@ttfamily{\textcolor{comcolor}{<!-- }}{\textcolor{comcolor}{<!-- }}}
    \lst@ProcessOther{"2D}{\lst@ttfamily{\textcolor{comcolor}{-->}}{\textcolor{comcolor}{-->}}}
    \@empty\z@\@empty
\makeatother

\lstdefinelanguage{XML}{
    morestring=[b]",
    morekeywords={action, typing, scroll, log, gaze, psi, level},
    morecomment=[s]{!}{-},
    basicstyle=\linespread{1.15}\footnotesize\ttfamily,
    backgroundcolor=\color{backcolor},
    keywordstyle=\color{keycolor},
    stringstyle=\color{valcolor},
    identifierstyle=\color{attrcolor},
    commentstyle=\color{comcolor},
}

\mdfdefinestyle{Quote}{
    backgroundcolor=gray!20, 
    linewidth=5pt, 
    linecolor=gray!50,
    rightline=false, 
    topline=false, 
    bottomline=false, 
}

\mdfdefinestyle{KeyFinding}{
    backgroundcolor=gray!20, 
    linewidth=2pt, 
    roundcorner=10pt,
}

\begin{document}

\title{Modeling Programmer Attention as\\Scanpath Prediction}
% \thanks{Identify applicable funding agency here. If none, delete this.}

%\author{\IEEEauthorblockN{Anonymous}
%\IEEEauthorblockA{
% \textit{Department of Computer Science and Engineering} \\
%\textit{~}\\
%~ \\
%~}
%}

%\author{\IEEEauthorblockN{Aakash Bansal, Chia-Yi Su, Zachary Karas, Yifan Zhang, Yu Huang, Toby Jia-Jun Li, and Collin McMillan}
%\IEEEauthorblockA{\textit{Dept. of Computer Science and Engineering} \\
%	\textit{University of Notre Dame}\\
%	Notre Dame, IN, USA \\
%	\{abansal1, csu3, cmc\}@nd.edu
%}}

\author{
    \IEEEauthorblockN{Aakash Bansal\IEEEauthorrefmark{1}, Chia-Yi Su\IEEEauthorrefmark{1}, Zachary Karas\IEEEauthorrefmark{2}, Yifan Zhang\IEEEauthorrefmark{2}, Yu Huang\IEEEauthorrefmark{2}, Toby Jia-Jun Li\IEEEauthorrefmark{1}, Collin McMillan\IEEEauthorrefmark{1}}
    \IEEEauthorblockA{\IEEEauthorrefmark{1}University of Notre Dame, USA
    \\\{abansal1,csu2,toby.j.li,cmc\}@nd.edu}
    \IEEEauthorblockA{\IEEEauthorrefmark{2}Vanderbilt University, USA
    \\\{z.karas, yifan.zhang.2, yu.huang\}@vanderbilt.edu}
}

\maketitle

\begin{abstract}
This paper launches a new effort at modeling programmer attention by predicting eye movement scanpaths. 
Programmer attention refers to what information people intake when performing programming tasks.  Models of programmer attention refer to machine prediction of what information is important to people. Models of programmer attention are important because they help researchers build better interfaces, assistive technologies, and more human-like AI.  For many years, researchers in SE have built these models based on features such as mouse clicks, key logging, and IDE interactions. Yet the holy grail in this area is scanpath prediction -- the prediction of the sequence of eye fixations a person would take over a visual stimulus.  A person's eye movements are considered the most concrete evidence that a person is taking in a piece of information.  Scanpath prediction is a notoriously difficult problem, but we believe that the emergence of lower-cost, higher-accuracy eye tracking equipment and better large language models of source code brings a solution within grasp.  We present an eye tracking experiment with 27 programmers and a prototype scanpath predictor to present preliminary results and obtain early community feedback.
\end{abstract}


\section{Introduction}
\label{sec:intro}

A machine model of human programmer attention is a computer's prediction of what information a person needs to solve a programming task.  Typically, these models input the source code of the software the programmer is developing, and attempt to predict what elements are most important for that programmer's understanding of the code.  These models are academically interesting on their own to improve our knowledge of human cognition~\cite{eivazi2011predicting}, but are also important building blocks to better interfaces~\cite{lu2016person}, assistive technologies for individuals with disabilities~\cite{majaranta2011gaze}, and even more human-like attention mechanisms in neural networks~\cite{lai2020understanding}.  Predicting human attention has been a core component of automated software engineering techniques for years~\cite{peitek2021program}.

Efforts to model programmer attention have trended towards predictions of actual human behaviors. Traditionally, researchers studied interactions such as mouse clicks, key strikes, or use of features in the programmers' Integrated Development Environments (IDE).  Yet over time, more studies have focused on synthetic visual attention (e.g., people selecting screen blurs with a mouse~\cite{paltenghi2021thinking}), eye fixations and tracking~\cite{bansal2023human}, and even live scans of brain activity~\cite{peitek2018simultaneous}.  The gold standard of modeling human attention is widely considered to be from the visual system, as a person's eye movements are considered the most concrete evidence available that a person is taking in a particular piece of information~\cite{lu2016person,hutmacher2019there}.  The trend in the literature is towards understanding these eye movements to understand what information a person needs.

Scanpath prediction refers to predicting the sequence of \textit{eye fixations} a person takes over a visual stimulus.  An eye fixation occurs when a person looks at a location of the stimulus long enough to intake the information at that location (on the order of 100ms~\cite{manor2003defining}). Scanpath prediction is a notoriously difficult problem because it depends on a person's thought process in addition to specific environmental factors~\cite{privitera2006scanpath}.  It may be easy to predict that a person will look at a bold, red word in a block of text, but it is harder to predict what word they will look at next to answer a particular intellectual question.  Therefore, scanpath prediction is emerging as a major research target~\cite{kummerer2021state}.

In this paper, we launch a new effort at modeling programmer attention via scanpath prediction.  Our new effort combines new eye tracking technologies, which allow more high-accuracy eye tracking experiments to be possible at lower cost, with new Large Language Models (LLMs) of source code, which have shown much better capabilities for code comprehension than previous generations.  We present:

\begin{description}[leftmargin=4mm]
\item[1.] An eye-tracking experiment with 27 programmers during a Java code comprehension task.  We ask programmers to read source code and write short summaries describing that code.  This task requires programmers to understand a snippet of code.  We use equipment that is available for $<$US\$10k and a web browser interface, which is a quarter of the price and much simpler setup than what was used in previous studies, allowing more data samples to be collected~\cite{rodeghero2014improving}.
\vspace{1mm}
\item[2.] A prototype scanpath predictor that is based on a 350m parameter language model of Java source code.  We frame the scanpath prediction problem as a fine-tuning objective in which the model receives the Java code as a prompt and the scanpath as a followup sequence of tokens to be predicted.  The idea is for the language model to learn to mimic the thought process of the programmers, insofar as the sequence of tokens that the programmers read.
\end{description}

Our goal is to present this paper as a paradigm for scanpath prediction of programmers, and obtain community feedback.

\newpage
\section{Related Work}

Techniques for scanpath prediction have been developed for decades. Initially, these consisted of either bio-inspired techniques that use neurophysical knowledge of the human visual system and processing ~\cite{ itti1998model,zanca2019gravitational,engbert2015spatial} or statistical models that used various distributions derived from a limited amount of data to model the eye gaze pattern~\cite{sun2014toward,clarke2017saccadic}. Since 2017, neural network based eye gaze models have offered an alternative to the more traditional statistical models. In particular, generative models like PathGAN~\cite{assens2018pathgan} and DeepGazeIII~\cite{kummerer2022deepgaze} have achieved state of the art performance. More recently, self-supervised techniques have been proposed that combine bio-inspired techniques and neural networks~\cite{schwinnbehind}. Although closely related to our problem, these models use saliency techniques specific to image processing that are often not transferable to textual data like source code.

Although eye-trackers have been intermittently used in automated SE research for over two decades, they have mainly been used as investigative tools to understand programmer cognition and behavior~\cite{sharif2012eye,sharafi2015systematic,sharafi2015eye,sharafi2021toward, tang2023empirical}. Models of programmer attention in software engineering research have typically used mouse cursors and keystrokes as proxy for programmer attention~\cite{huber2023look,liu2018predicting}. Earlier this year, Bansal~\emph{et al.}~\cite{bansal2023human} introduced a novel approach for predicting human attention using data from a 2014 eye-tracking study~\cite{rodeghero2014improving}. They introduced a model to predict programmer attention, in the form of total fixation duration. They cite the unavailability of recent eye-tracking data as a limitation for their approach. Therefore, in this paper we conduct a larger eye-tracking study designed to inform an attention prediction model. To the best of our knowledge, this paper is only the second attempt at modeling programmer attention using eye trackers, and the first to provide a prototype for scanpath prediction which is a more challenging problem.

\vspace{-0.4cm}
\section{Eye-Tracking Experiment}
\label{sec:eyetracking}
We perform an eye-tracking experiment to extract the scanpath data of programmers for modeling programmer attention. Figure~\ref{fig:overview} shows an overview of our experiments: 1) this eye-tracking experiment, 2) the prototype in Section~\ref{sec:attendex}, and 3) preliminary experiment in Section~\ref{sec:attentionexp}.

\vspace{-0.1cm}
\subsection{Program Comprehension Task}
\vspace{-0.05cm}
To record eye gaze data and extract the scanpath, we recruited programmers for a program comprehension task, specifically, code summarization. Code summarization is the task of writing natural language summary for a snippet of code, a Java method in this instance. We chose this task, because it requires the programmer to bridge the gap between the lower level understanding of source code and the high-level concepts that benefit from human cognition and attention~\cite{biggerstaff1994program}. 

We populate our study using the {\small\texttt{funcom-java-long}} dataset~\cite{bansal2023human}. The dataset consists of high-quality Java method-summary pairs from Javadocs. This dataset is also excluded from the {\small\texttt{jm52m}}~\cite{jm52m} dataset that was used for pre-training of our prototype, detailed in Section~\ref{sec:attendex}. From the roughly 8800 Java methods in their test set, we picked 68 methods at random for our experiment. Each participant saw 25 randomly selected methods from the subset.

\begin{figure}[t]
	\centering
	\includegraphics[width=0.99\linewidth]{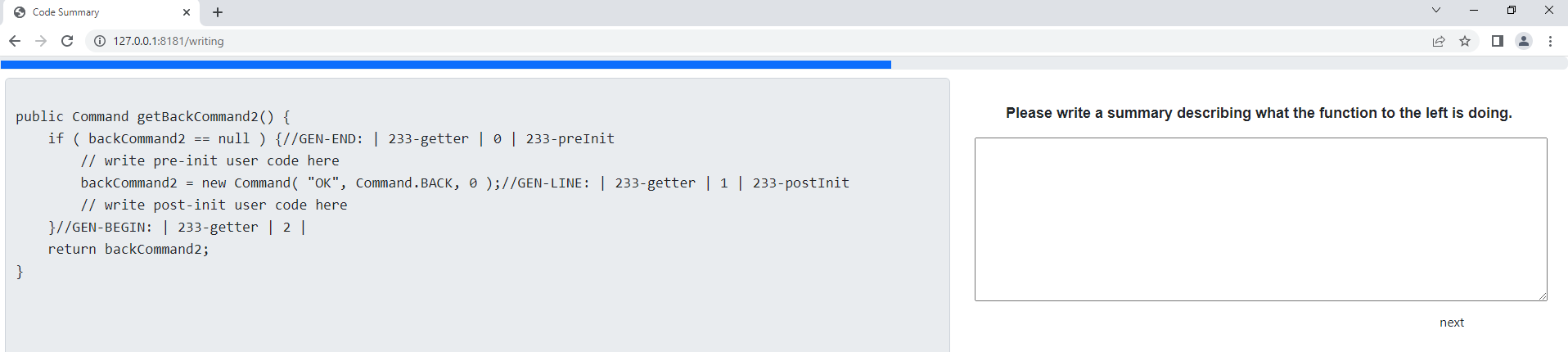}
	\vspace{-0.6cm}
    \caption{A screenshot of our interface.}
    \label{fig:screenshot}
    \vspace{-0.6cm}
\end{figure}

In Figure~\ref{fig:screenshot} we show an example of the web interface and the task we asked each participant to complete. Each task consists of a Java method displayed on the left, and a text box on the right. For this study, we recruited 27 programmers, each with 3-10 years of programming experience, and a minimum of 1 year of Java programming experience.
We asked each participant to complete 25 tasks, with an average study duration of 1.5 hours. We asked participants to take a break every 20 minutes to re-calibrate the eye-tracker and minimize the effects of exhaustion ~\cite{sharafi2015eye}.

 %We compensated each programmer at \$66 per hour. 

\vspace{-0.1cm}
\subsection{The Eye-Tracking Hardware \& Software}
\vspace{-0.1cm}

The eye-tracking hardware we use includes a Tobii Pro Fusion eye-tracker and a 24 inch monitor at 1920x1080 resolution and a refresh rate of 60Hz. We kept the workstation costs below \$10K, much lower than \$40K in previous studies~\cite{rodeghero2014improving}. Lower costs and the mobile nature of the eye-tracker setup allowed us to collect more data at two locations. The eye-tracker is mounted on the monitor, which is beneficial because the participants did not have to change their work-flow and were largely left to work as they would without the eye-tracker. We use the Tobii Pro Python SDK to access raw gaze data. 

\vspace{-0.15cm}
\subsection{Data Collection \& Processing}
\label{sec:eyedata}
\vspace{-0.05cm}

\begin{figure}[t]
	\centering
	\hspace{-0.25cm}\includegraphics[width=1.01\columnwidth]{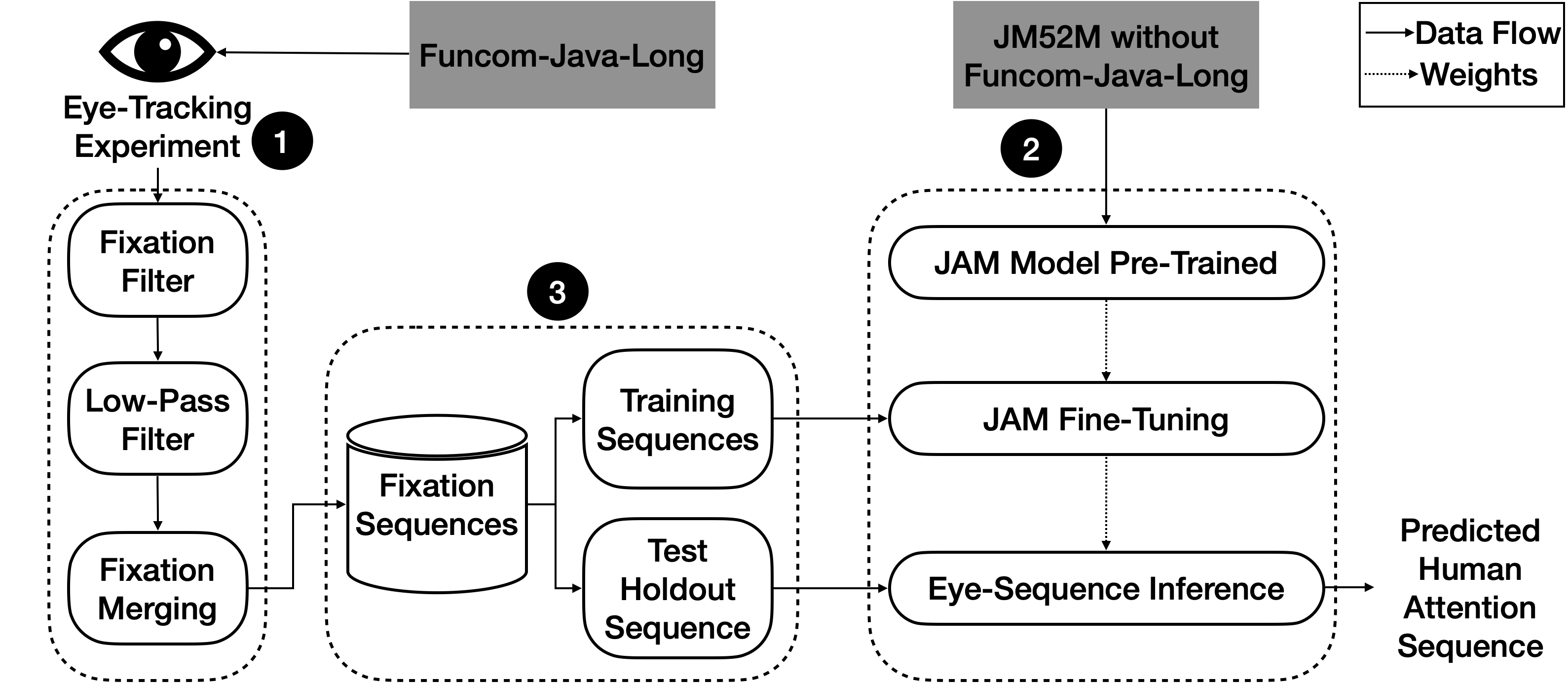}
	\vspace{-0.2cm}
    \caption{Overview of our experimental setup.}
    \label{fig:overview}
	\vspace{-0.3cm}
\end{figure}

In Figure~\ref{fig:overview} area 1, we provide an overview of the eye-tracker data processing. We performed this study at two institutions in parallel (anonymized for review). We used the same data set, interface, hardware, and protocol for both studies. First we apply a velocity-based fixation filter, then we use a low-pass filter to remove noisy peaks~\cite{olsen2012tobii,mack2017effect}. Next, we merge the data from both institutions. The only difference between the two setups is that one eye tracker collected the data at a sampling rate of 120Hz, while the other collected data at 60Hz due to software mismatch. This difference in sampling rates does not affect our study as we cluster fixations using methods recommended in related work for eye-tracking data with different sampling frequencies~\cite{ouzts2012comparison,trabulsi2021optimizing}.The final result of this processing is 680 data points, each containing a sequences of first $n$ fixations from a participant over a method.

\section{Scanpath Predictor Prototype}
\label{sec:attendex}

\begin{figure}[t!]

\begin{scriptsize}
\begin{verbatim}
TDAT:   public void  testNegativeParseCases() {
    verbose("--->Negative parse tests  START");
    for (int i = 0; i < negativeParseTests.length; i++) {
      parseFilter(negativeParseTests[i], false);
    }
    checkDelete(); }
 SEQ: <s> testNegativeParseCases </s>
\end{verbatim}
\end{scriptsize}
\normalsize
\vspace{-0.4cm}
\caption{Finetuning prompt for the participant ID 133, method ID 31696447.}
\vspace{-0.6cm}
\label{fig:prompt}
\end{figure}

We design a prototype scanpath predictor to model programmer attention using the scanpath data we extract from the eye-tracking experiment. We provide an overview of our design in Figure~\ref{fig:overview} area 2, where we frame the problem as a fine-tuning task for an LLM. We use LLMs because they are pre-trained on large task-agnostic datasets and can be finetuned for a prediction task given a small number of examples. Although our eye-tracking apparatus is much cheaper than previous generation of eye-trackers, recruitment of programmers for the study is still a cost-limiting factor. Therefore, LLMs are good candidates for our prototype. For our prototype we use the {\small\texttt{jam}}~\cite{su2023language} language model. We chose this specific pre-trained LLM for two reasons. First, at $\sim$ 350m parameters the model is large enough to produce meaningful results but small enough to fit on a single 24GB GPU~\cite{su2023language}. Second, the model is pre-trained on the {\small\texttt{jm52m}}~\cite{jm52m} dataset of 52 million Java methods which excludes the {\small\texttt{funcom-java-long}} dataset that we use for our eye-tracking experiment. Therefore, by using {\small\texttt{jam}}, we can be reasonably sure that the Java methods we use for our experiment and subsequent testing, have not been previously seen by the model during pre-training. 

We finetune our prototype on the training set extracted from the eye-tracking experiment. During finetuning we provide the model with a prompt which consists of raw Java code, followed by the scanpath sequence. Figure~\ref{fig:prompt} shows an example of our finetuning prompt. We use the following configurations to finetune our model:
\vspace{-0.1cm}
\begin{table}[h!]
    \centering
	\footnotesize
	\begin{tabular}{p{1cm}p{3.5cm}p{1cm}}
		$c$ & block size                        & 256  		 \\
		$b$ & batch size                        & 4  		 \\
		$e$ & embedding dimension               & 768 		 \\
		$L$ & number of layers 			  		& 12  		 \\
		$h$ & attention heads             		& 12 		 \\
		$a$ & accumulation steps				& 32		 \\
		$r$ & learning rate                     & 3e-5 		 \\
        $s$ & pre-trained iterations		    & 27200	     \\
		$i$ & iterations for finetuning			& 200		 \\
	\end{tabular}
 \vspace{-0.35cm}
\end{table}

After finetuning, we provide the model with the test set for inference. The input prompt is similar to the prommpt that we use for finetuning in Figure~\ref{fig:prompt}, except that we stop the prompt at ``SEQ:''. The prototype predicts {\small\texttt{<s> {scanpath} </s>}}, where {\small\texttt{\{scanpath\}}} is a sequence of $n$ predictions, and {\small\texttt{<s>}} and  {\small\texttt{</s>}} are start and end tags for the sequence.

\section{Preliminary Experiment}
\label{sec:attentionexp}
We conduct a preliminary experiment to evaluate the efficacy of our scanpath predictor prototype. Recall, we design our prototype to predict first $n$ words in the scanpath. For this experiment we evaluate predictions at $n=\{1,2,3,4\}$.

Before we dive into the experimental details, we introduce the research questions for this experiment:

\begin{description}
\item[\textbf{RQ1}]  How accurately can our prototype predict the first $n$ words that a programmer would look at when summarizing a Java method previously seen during finetuning?

\item[\textbf{RQ2}]  How accurately can our prototype predict the first $n$ words that programmers would look at when summarizing a previously unseen Java method?
\end{description}

The rationale behind RQ1 is to evaluate the accuracy of the scanpath predictor prototype against reference data from a particular programmer. Note, the model has seen the scanpath of the other participants over the same Java method during finetuning. The goal is to evaluate the correlation between the predicted scanpath and a human programmer's scanpath.

The rationale behind RQ2 is to challenge the scanpath predictor prototype. Note, the model has never seen this Java method during pre-training or finetuning. The goal is to evaluate how accurately the prototype predicts the scanpath over an unseen method, compared to the human programmers.

\vspace{-0.2cm}
\subsection{Holdout Experiment Setup}

In Figure~\ref{fig:overview} area 3, we provide an overview of our preliminary experiment. Our goal is to compare the predictions from our prototype against reference scanpath from human subjects. However, different programmers might look at different words in the sequence. Therefore, we do not test our approach against any one programmer's data. Instead, we use a one-holdout approach and create 95 versions of our dataset. 

For RQ1, we take a {\small\texttt{participant-holdout}} approach, where we holdout all the datapoints from one participant as the test set. The training data in each instance has 26 participants, of which 1 participant is held out for validation. The test data in each instance has scanpaths from 1 participant over each of the 25 methods they saw during the experiment.

For RQ2, we take a {\small\texttt{method-holdout}} approach, where we holdout all the datapoints from for one Java method as the test set. The training data in each instance has 67 methods, of which 1 method is held out for validation during finetuning. The test data in each instance has scanpaths from 1 method over a variable number of participants because not every participant saw every method. Recall from Section~\ref{sec:eyetracking}, we have 27 participants and 68 methods. Therefore, we processed 27 datasets for the {\small\texttt{participant-holdout}} experiments and  68 datasets for the {\small\texttt{method-holdout}} experiment. 

\vspace{-0.2cm}
\subsection{Metrics}
We report two metrics from related work to evaluate the performance of our prototype scanpath predictor.

\begin{description}
\item[Levenshtein] is the string-matching metric that uses the Levenshtein edit distance, which is a popular scanpath metric~\cite{fahimi2021metrics}, to compute character level string comparison. We use the TheFuzz~\cite{fuzzymatch} implementation.

\item[Gestalt] is a popular pattern-matching metric that uses the Gestalt Pattern Matching algorithm to compute the similarity between two sequences. We use the pyymatcher~\cite{pyymatcher} implementation.
\end{description}
\setlength{\tabcolsep}{1.3pt}

\begin{table}[b!]
\centering
 \vspace{-0.6cm}
\caption{Average Gestalt and Levenshtein scores for RQ1 and RQ2.}
\label{tab:roc}
\begin{tabular}{l|llll|llll}
        Experiment  & \multicolumn{4}{|c|}{Levenshtein} & \multicolumn{4}{|c}{Gestalt}\\ 
		    & $n=1$ & $n=2$   & $n=3$  &  $n=4$ & $n=1$ & $n=2$   & $n=3$  &  $n=4$\\ \cline{1-9}
		participant-holdout  & 0.431      & 0.434      &  0.452     & \textbf{0.460}     &\textbf{0.442}       & 0.421       & 0.429      & 0.424 \\
		method-holdout      & 0.337      & 0.342     &   \textbf{0.343}    & 0.332   &\textbf{0.347}      & 0.332      & 0.315      & 0.295 \\ 
	\end{tabular}

\end{table}

\section{Results for Preliminary Experiment}
We present the results for both of our preliminary experiments to evaluate our prototype scanpath predictor. In Table~\ref{tab:roc} we show the average scores over for each experiment. 

%In the next two subsections, we discuss the scores and provide a more in-depth analysis.

\vspace{-0.15cm}
\subsection{RQ1: Participant Holdout Experiment}

For the {\small\texttt{participant-holdout}} experiment, we observe  that the Levenshtein score is the highest at 0.46 when $n=4$, which means that the prototype is slightly better at predicting longer scanpaths than shorter $n<=3$. This is a promising result because we expect future research to focus on predicting scanpaths at $n>4$. On the other hand, for the Gestalt metric, the highest score is 0.442 at $n=1$. We observe that the Levenshtein scores show a consistent increase from $n=1-4$, while the Gestalt metric is less comparative between $n=2-4$. Each of the scores in Table~\ref{tab:roc} is an average over 680 prediction-reference pairs, therefore we dig a little deeper into the score distribution to understand these scores.

In Figure~\ref{fig:fuzzy} (a), we show a frequency distribution bar-chart of our Levenshtein scores. We observe that for $n=1$, the highest frequency is at the Levenshtein score of 1.0, which means a perfect match for 168 out of 680 samples. This is a promising result for our prototype, because it shows that it learns to pick the first word correctly out of all the words in the Java method for 25\% of the samples. For $n=2,3,4$ we observe a sharp decrease in perfect matches with score 1, but a considerable increase in number of samples with scores in the $[0.4-0.8]$ range, which is above the average scores we saw in Table~\ref{tab:roc}. Overall, we find that although our prototype achieves higher average scores for $n=4$, the number of perfect matches are very few ($<5$). One possible explanation is that scanpaths vary between participants and neighboring words in a code sequence may not match the target word, making a perfect match hard to achieve for longer scanpaths.

\vspace{-0.1cm}
\subsection{RQ2: Method Holdout Experiment}

For the {\small\texttt{method-holdout}} experiment, we observe a highest Levenshtein scores are less comparative for $n=1-4$ with the highest score of $0.343$ for $n=3$. The Gestalt score is highest for $n=1$, and is more comparative with a consistent trend of decreasing values for $n=1-4$. Overall, we see that the average scores are lower compared to RQ1. We expect this because predicting scanpath over an unseen method is a harder task. The model did not see this method during pre-training, making it more challenging to learn word embeddings for unique words such as variable names.

\begin{figure}[t!]

	\centering
	\includegraphics[width=0.5\textwidth]{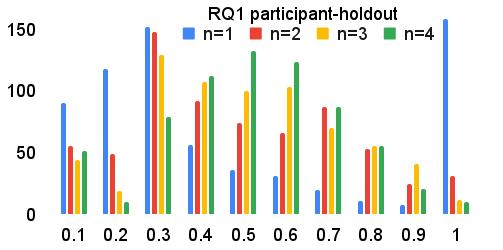}\\
    \vspace{-0.15cm}
    (a)\\
    \includegraphics[width=0.5\textwidth]{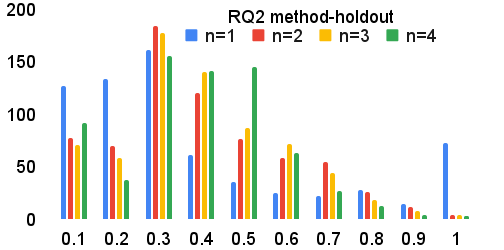}\\
    \vspace{-0.15cm}
    (b)\\
    \vspace{-0.15cm}
    \caption{\raggedright Frequency charts for (a) \texttt{participant-holdout} and (b) \texttt{method-holdout} experiments. There are 680 samples. The x axis indicates bins of Levenshtein score, the y axis indicates number of samples.}
    \label{fig:fuzzy}
    \vspace{-0.65cm}
\end{figure}

To investigate further, we show a frequency distribution bar-chart of Levenshtein scores in Figure~\ref{fig:fuzzy} (b). We note that for $n=2,3,4$ we observe a sharp decrease in the number of perfect matches when compared to $n=1$, and a significant increase in number of samples with scores in the $[0.4-0.8]$ range. This aligns with our expectations, in that predicting scanpaths over unseen methods is a harder task, specially for longer scanpaths. We expect future work with bigger eye-tracking datasets to improve these results because our prototype may benefit from seeing scanpaths of similar methods during finetuning and learn from them. 

\vspace{-0.05cm}
\section{Conclusion}
\label{sec:conclusion}
In this new ideas paper we make three main contributions to model programmer attention. First, we present an eye-tracking experiment designed to extract scanpath data from programmers. Second, we frame the scanpath prediction problem as a finetuning task for an LLM to develop a novel prototype. Third, we show how well our prototype correlates with human scanpaths with a preliminary experiment. The main goal of this paper is to provide a framework for future research towards modeling programmer attention, specifically by automatically predicting eye movements such as scanpaths. Potential applications of this work are towards building more human-like neural networks~\cite{peters2021capturing}, SE virtual assistants that consider human factors~\cite{eberhart2020wizard}, and assisted systems for low vision and disabled programmers~\cite{armaly2018comparison,potluri2019ai} to name a few. 

We provide a repository for the replication of our results at:

\url{https://github.com/apcl-research/scanpathpred}

\vspace{-0.1cm}
\section*{Acknowledgment}
This work is supported in part by the NSF CCF-2100035 and CCF-2211428. Any opinions, findings, and conclusions expressed herein are the authors’ and do not necessarily reflect those of the sponsors.

\clearpage
\bibliographystyle{IEEEtran}
\bibliography{main}

\end{document}